\documentclass [11pt,a4paper]{article}
\usepackage{amsmath,amssymb,amsbsy,amsfonts,amsthm,latexsym,amsopn,amstext,                                                amsxtra,euscript,amscd}
\def\F{\mathbb{F}}

\def\Z{\mathbb{Z}}

\usepackage{color,xcolor}
\usepackage{float}
\usepackage{graphicx}
\newtheorem{theorem}{Theorem}
\newtheorem{lemma}{Lemma}
\newtheorem{corollary}{Corollary}

\newcommand{\quash}[1]{}

\begin{document}

\title{On the $k$-error linear complexity of binary
sequences derived from polynomial quotients}

\author{Zhixiong Chen\\
School of Applied Mathematics, Putian University, \\ Putian, Fujian
351100, P. R. China\\
ptczx@126.com\\
\\
Zhihua Niu\\
School of Computer Engineering and Science, Shanghai University,\\
Shangda Road, Shanghai 200444, P. R. China \\
zhniu@staff.shu.edu.cn\\
\\
Chenhuang Wu\\
School of Applied Mathematics, Putian University, \\ Putian, Fujian
351100, P. R. China}
\maketitle

\begin{abstract}
We investigate the $k$-error linear complexity of $p^2$-periodic
binary sequences defined from the polynomial quotients (including the well-studied Fermat quotients), which is defined by
$$
q_{p,w}(u)\equiv \frac{u^w-u^{wp}}{p} \bmod p ~~  \mathrm{with}~~ 0
\le q_{p,w}(u) \le p-1, ~~u\ge 0,
$$
where $p$ is an odd prime and $1\le w<p$. Indeed, first for all integers $k$, we determine exact values of the $k$-error linear complexity over the finite field $\F_2$ for these binary sequences under the assumption of $2$ being a primitive root modulo $p^2$, and then we determine their $k$-error linear complexity over the finite field $\F_p$ for either $0\le k<p$ when $w=1$ or $0\le k<p-1$ when $2\le w<p$. Theoretical results obtained indicate that such sequences possess `good' error linear complexity.
\end{abstract}



\section{Introduction}\label{intro}

For an odd prime $p$ and integers $u\ge 0$ with $\gcd(u,p)=1$, the {\it
Fermat quotient $q_p(u)$ \/} is defined as the unique integer
$$
q_p(u) \equiv \frac{u^{p-1} -1}{p} \bmod p ~~  \mathrm{with}~~ 0 \le
q_p(u) \le p-1,
$$
and
$$
q_p(lp) = 0, \qquad l \in \mathbb{Z}.
$$
An equivalent definition of the Fermat quotient is given below
\begin{equation}\label{FFF}
q_p(u)\equiv \frac{u^{p-1}-u^{p(p-1)}}{p}\bmod p, ~~ u\ge 0.
\end{equation}
For any fixed positive integer $w$, by the fact that
$$
(u^w)^{p}\equiv u^w \bmod p, ~~ u\ge 0
$$
from the Fermat Little Theorem, Chen and Winterhof extended (\ref{FFF})
to define
\begin{equation}\label{poly-def}
q_{p,w}(u)\equiv \frac{u^w-u^{wp}}{p} \bmod p ~~  \mathrm{with}~~ 0
\le q_{p,w}(u) \le p-1, ~~u\ge 0,
\end{equation}
which is called a \emph{polynomial quotient} in \cite{CW2}. In
fact $q_{p,p-1}(u)=q_{p}(u)$. It is easy to see that
\begin{equation}\label{addstruct}
q_{p,w}(u+lp) = q_{p,w}(u) + wl u^{w-1}~( \bmod~ p)
\end{equation}
if $\gcd(u,p)=1$, and
\begin{equation}\label{value-2}
q_{p,w}(lp) =\left\{
\begin{array}{ll}
0, & \mathrm{if}\,\, w>1,  \\
l, & \mathrm{if}\,\, w=1,  \\
\end{array}
\right. ~l=0,\ldots,p-1.
\end{equation}

Many number theoretic and cryptographic questions as well as
measures of pseudorandomness have been studied for Fermat quotients
and their generalizations
\cite{ADS,AW,BFKS,C,CD,CG,COW,CW2,CW,CW3,DCH,DKC,EM,GW,OS,Sha,Shk,S,S2010,S2011,S2011b,SW}.

In this paper, we still concentrate on certain binary sequences defined from the polynomial quotients (of course including the Fermat quotients) in the references.
The first one is the binary threshold sequence $(e_u)$ studied in \cite{CD,CG,CHD,COW,CW4,DKC} by defining
\begin{equation}\label{binarythreshold}
e_u=\left\{
\begin{array}{ll}
0, & \mathrm{if}\,\ 0\leq q_{p,w}(u)/p< \frac{1}{2},\\
1, & \mathrm{if}\,\ \frac{1}{2}\leq q_{p,w}(u)/p< 1,
\end{array}
\right. \quad  u \ge 0.
\end{equation}
The second one, by combining $q_{p,w}(u)$ with the Legendre symbol
$\left( \frac{\cdot}{p} \right)$, is defined in \cite{CHD,CW4,DKC,GW} by
\begin{equation}\label{binarylegendre}
f_u=\left\{
\begin{array}{ll}
0, & \mathrm{if}\,\ \left(\frac{q_{p,w}(u)}{p}\right)=1\,\ \mathrm{or}\,\ q_{p,w}(u)=0, \\
1, & \mathrm{otherwise},
\end{array}
\right. \quad u \ge 0.
\end{equation}
(In fact, in \cite{DKC,GW} $\chi$, a fixed multiplicative
character modulo $p$ of order $m>1$, is applied to defining $m$-ary sequences $(\widetilde{f}_u)$ of
discrete logarithms modulo a divisor $m$ of $p-1$ by
$$
  \exp(2\pi i \widetilde{f}_u/m)=
     \chi(q_{p,w}(u)),~ 0\le \widetilde{f}_u<m \quad \mbox{if } q_{p,w}(u)\not\equiv 0\bmod p
$$
and $\widetilde{f}_u=0$ otherwise. When $m=2$, we have $\widetilde{f}_u=f_u$ for all $u\ge 0$.) We note that both $(e_u)$ and $(f_u)$ are $p^2$-periodic by (\ref{addstruct}).

The authors of \cite{COW,GW} investigated measures of pseudorandomness as well as linear complexity profile of
$(e_u)$ and $(\widetilde{f}_u)$ (of course including $(f_u)$) via certain character sums over Fermat quotients.
The authors of \cite{CHD,DKC}  determined the \emph{linear complexity} (see
 below for the definition) of $(e_u)$ and $(f_u)$ if $2$ is a primitive element modulo $p^2$, and later the authors of \cite{CD,CG,CW4}  extended to a more general setting of $2^{p-1}\not\equiv 1 \pmod {p^2}$ when $w\in \{p-1, (p-1)/2\}$. The authors of \cite{CW4}  also determined the  trace representations of $(e_u)$ and $(f_u)$. In this paper, our main aim is to study the \emph{$k$-error linear complexity} (see  below for the definition) for $(e_u)$ and $(f_u)$.
All results indicate that such sequences have desirable cryptographic features.

For our purpose, we need to describe $(e_u)$  and $(f_u)$ in an equivalent way. From (\ref{addstruct}),
$q_{p,w}(-)$ induces a surjective map from $\Z_{p^2}^*$ (the group of invertible
elements modulo $p^2$) to $\Z_p$ (the additive group of numbers modulo $p$). For each fixed $1\le w<p$, we define
$$
D_l=\{u: 0\le u< p^2,~ \gcd(u,p)=1,~q_{p,w}(u)=l\}
$$
for $l=0,1,\ldots,p-1$. Each $D_l$ has the cardinality $|D_l|=p-1$ by (\ref{addstruct}). Here and hereafter, we use $|S|$ to denote the cardinality of a set $S$. Let $P=\{lp: 0\le l< p\}$, for $w\ge 2$
one can define $(e_u)$  and $(f_u)$ equivalently by
$$
e_u=\left\{
\begin{array}{ll}
0, & \mathrm{if}\,\ u \bmod {p^2}\in D_0 \cup \cdots \cup D_{(p-1)/2} \cup P,\\
1, & \mathrm{if}\,\ u\bmod {p^2} \in D_{(p+1)/2} \cup \cdots \cup D_{p-1},
\end{array}
\right.
$$
and
$$
f_u=\left\{
\begin{array}{ll}
0, & \mathrm{if}\,\ u\bmod {p^2}\in \cup_{l\in Q} D_l  \cup D_0\cup P,\\
1, & \mathrm{if}\,\ u\bmod {p^2}\in \cup_{l\in N} D_{l},
\end{array}
\right.
$$
respectively, where $Q$ is the set of quadratic residues modulo $p$ and $N$ is the set of quadratic non-residues modulo $p$.
For $w=1$, it is easy to  define $(e_u)$  and $(f_u)$ similarly by only re-dividing the set $P$.

We need to mention that,  the following relation holds  between
$q_{p,w}(u)$ and $q_{p}(u)$:
\begin{equation}\label{poly-Fermat:relation}
q_{p,w}(u)\equiv -u^wwq_{p}(u) \bmod p
\end{equation}
for all $u\ge 0$ with $\gcd(u,p)=1$.
If $w=lp$ for any positive integer $l$, we have $q_{p,lp}(u)=0$ by
(\ref{poly-Fermat:relation}) and (\ref{value-2})  for all $u\ge 0$.
For any positive $w$ with $p\nmid w$, write $w=w_1+w_2(p-1)$ with
$1\le w_1\le p-1$ and $w_2\ge 0$, by (\ref{poly-Fermat:relation})
again one can get
$$
q_{p,w_1+w_2(p-1)}(u)\equiv -u^{w_1}(w_1-w_2)q_{p}(u)\equiv
w_1^{-1}(w_1-w_2)q_{p,w_1}(u) \bmod p.
$$
Note that $w_1\not\equiv w_2
\bmod p$ since  $p\nmid w$. Hence, a polynomial quotient $q_{p,w}(-)$ with large $w$ can be reduced to the one with $1\le w_1\le p-1$ and we restrict ourselves to
$1\le w\le p-1$ from now on.

We conclude this section by introducing the notions of the linear complexity and the $k$-error linear complexity of periodic sequences.

Let $\F$ be a field.  For a $T$-periodic
sequence $(s_u)$ over $\F$, we recall that the
\emph{linear complexity} over $\F$, denoted by  $LC^{\F}((s_u))$, is the least order $L$ of a linear
recurrence relation over $\mathbb{F}$
$$
s_{u+L} = c_{L-1}s_{u+L-1} +\cdots +c_1s_{u+1}+ c_0s_u\quad
\mathrm{for}\,\ u \geq 0,
$$
which is satisfied by $(s_u)$ and where $c_0\neq 0, c_1, \ldots,
c_{L-1}\in \mathbb{F}$.
Let
$$
S(X)=s_0+s_1X+s_2X^2+\cdots+s_{T-1}X^{T-1}\in \mathbb{F}[X],
$$
which is called the \emph{generating polynomial} of $(s_u)$. Then the linear
complexity over $\F$ of $(s_u)$ is computed by
\begin{equation}\label{licom}
  LC^{\F}((s_u)) =T-\deg\left(\mathrm{gcd}(X^T-1,
  ~S(X))\right),
\end{equation}
see, e.g. \cite{LN} for details. For integers $k\ge 0$, the \emph{$k$-error linear complexity} over $\F$ of $(s_u)$, denoted by $LC^{\F}_k((s_u))$, is the smallest linear complexity (over $\F$) that can be
obtained by changing at most $k$ terms of the sequence per period, see \cite{SM,Meidl}, and see \cite{DXS} for the related even earlier defined sphere complexity.  Clearly $LC^{\F}_0((s_u))=LC^{\F}((s_u))$ and
$$
T\ge LC^{\F}_0((s_u))\ge LC^{\F}_1((s_u))\ge \ldots \ge LC^{\F}_k((s_u))=0
$$
when $k$ equals the number of nonzero terms of $(s_u)$ per period, i.e., the weight of $(s_u)$.

The linear complexity and the $k$-error linear complexity are important cryptographic characteristics of sequences
and provide information on the predictability and thus unsuitability for cryptography. For a sequence to be cryptographically strong, its linear complexity
should be large, but not significantly reduced by changing a few
terms. And according to  the Berlekamp-Massey
algorithm \cite{Massey}, the linear complexity
should be at least a half of the period.

Instead of studying $(e_u)$  and $(f_u)$ directly, we define the $p^2$-periodic binary sequence $(h_u)$ by
\begin{equation}\label{hhhh}
h_u=\left\{
\begin{array}{ll}
1, & \mathrm{if}\,\ u\bmod {p^2}\in \cup_{l\in \mathcal{I}} D_l,\\
0, & \mathrm{otherwise},
\end{array}
\right. \quad u\ge 0,
\end{equation}
for $w\ge 2$, and
\begin{equation}\label{hhhh-w=1}
h_u=\left\{
\begin{array}{ll}
1, & \mathrm{if}\,\ u\bmod {p^2}\in \cup_{l\in \mathcal{I}} (D_l\cup \{lp\}),\\
0, & \mathrm{otherwise},
\end{array}
\right. \quad u\ge 0,
\end{equation}
for $w=1$, where $\mathcal{I}$ is a non-empty subset of $\{0,1,\ldots,p-1\}$, and investigate the $k$-error linear complexity over $\F_2$ for $(h_u)$ in Section \ref{LC-2}. In Section \ref{LC-p}, we investigate the $k$-error linear complexity over $\F_p$ for $(h_u)$. Although $(h_u)$ is a binary sequence, it is constructed based on the polynomial quotients modulo $p$ (note that the linear complexity over $\F_p$ of the polynomial quotients is $p+w$, see a proof in  \cite{OS} for the Fermat quotients), thus, it is natural to consider the $k$-error linear complexity over $\F_p$ for $(h_u)$. In fact, it is also motivated by the ideas of \cite{AMW,AW06} and partially \cite{AM,AW,BW,CY,ESK,GLSW,HKN,HMMS}.

\section{$k$-Error Linear Complexity over $\F_2$}\label{LC-2}

First we present some auxiliary statements. Define
$$ D_l(X)= \sum\limits_{u\in D_l}X^u \in \mathbb{F}_2[X]$$
for $0\leq l < p$.

\begin{lemma}\label{lemma-add}
Let $\theta \in \overline{\mathbb{F}}_{2}$ be  a  primitive $p$-th root of
unity. For $0\leq l < p$, we have
$$
D_l(\theta^{m})=\left\{
\begin{array}{ll}
0, & \mathrm{if}~ m\equiv 0 \pmod p,\\
1, & \mathrm{otherwise}.
\end{array}
\right.
$$
\end{lemma}
Proof. For any fixed $1\le v<p$, the numbers $v+mp$ belong to different $D_l~ (0\leq l < p)$ when $m$ runs through the set $\{0,1,\ldots,p-1\}$ by  (\ref{addstruct}), hence we have
$$
\{u\pmod p: u\in D_l\}=\Z_p^*, ~~~0\leq l < p.
$$
 We note that in the definition of $D_l$, we restrict $1\le w<p$. For $0\leq l < p$, we derive
$$
 D_l(\theta^{m})=\sum\limits_{u\in D_l}\theta^{mu}=\sum\limits_{j\in \Z_p^*}\theta^{mj},
 $$
which deduces the desired result for different $m$ modulo $p$. The calculations here are performed in
finite fields with characteristic two. ~\hfill $\square$

\begin{lemma}\label{lemma-p}
Let $\theta \in \overline{\mathbb{F}}_{2}$ be  a  primitive $p$-th root of
unity and $G(X)\in \F_2[X]$ with $1\le \deg(G(X))<p$. If  $2$ is a
primitive root modulo $p$, we have
$$
G(\theta)=1  \Longleftrightarrow G(X)=X+X^2+\ldots+X^{p-1},
$$
or
$$
G(\theta)=0  \Longleftrightarrow G(X)=1+X+X^2+\ldots+X^{p-1}.
$$
\end{lemma}
Proof. We only show the first assertion. Since $2$ is a
primitive root modulo $p$, we see that $1+X+X^2+\ldots+X^{p-1}$ is the minimal irreducible polynomial with the root $\theta$.
So if $G(\theta)=1$, we derive
$$
(1+X+X^2+\ldots+X^{p-1}) | (G(X)-1).
$$
With the restriction on $\deg(G(X))$, we get $G(X)=X+X^2+\ldots+X^{p-1}$. The converse is true after simple calculations.
~\hfill $\square$

Now we present our main results.

\begin{theorem}\label{klc-2-primitive}
Let $(h_u)$ be the binary sequence of period $p^{2}$ defined in (\ref{hhhh}) using polynomial quotients (\ref{poly-def}) with $2\le w\le p-1$ and a non-empty subset $\mathcal{I}$ of $\{0,1,\ldots,p-1\}$ with $1\le |\mathcal{I}|\le (p-1)/2$. If $2$ is a primitive root modulo $p^2$, then
the $k$-error linear complexity  over $\F_2$ of $(h_u)$  satisfies
\[
 LC^{\F_2}_k((h_u))=\left\{
\begin{array}{cl}
p^2-1, & \mathrm{if}\,\ k=0, \\
p^2-p+1, & \mathrm{if}\,\ 1\le k<p-1, \\
p^2-p, & \mathrm{if}\,\ p-1\le k< (p-1)|\mathcal{I}|, |\mathcal{I}|>1, \\
0, & \mathrm{if}\,\ k\ge (p-1)|\mathcal{I}|, \\
\end{array}
\right.\\
\]
if $|\mathcal{I}|$ is odd, and otherwise
\[
 LC^{\F_2}_k((h_u))=\left\{
\begin{array}{cl}
p^2-p, & \mathrm{if}\,\ 0\le k<(p-1)|\mathcal{I}|, \\
0, & \mathrm{if}\,\ k\ge (p-1)|\mathcal{I}|.
\end{array}
\right.
\]
\end{theorem}
Proof. From the construction of $(h_u)$, there are $(p-1)|\mathcal{I}|$ many 1's in one period of $(h_u)$ since each $D_l$ contains $p-1$ many elements. Changing
all terms of 1's will lead to the zero sequence. So we always assume that $k<(p-1)|\mathcal{I}|$. Let
\begin{equation}\label{Hk}
H_k(X)=\sum\limits_{l\in \mathcal{I}}D_{l}(X)+e(X)\in \F_2[X]
\end{equation}
be the generating polynomial of the sequence obtained from $(h_u)$ by changing exactly $k$ terms of $(h_u)$ per period,
where $e(X)$ is the corresponding error polynomial with $k$ many  monomials. We note that $H_k(X)$ is a nonzero polynomial due to $k<(p-1)|\mathcal{I}|$. We will consider the common roots of $H_k(X)$ and $X^{p^2}-1$,
i.e.,  the roots of the form $\beta^{n}  ~ (n\in \Z_{p^2})$ for $H_k(X)$, where $\beta \in \overline{\mathbb{F}}_{2}$ is a  primitive $p^2$-th root of unity. The number of the common roots will help us to derive the values of $k$-error linear
complexity of $(h_u)$  by  (\ref{licom}).

On the one hand, we assume that $H_k(\beta^{n_0})=0$ for some $n_0\in \Z_{p^2}^*$. Since $2$ is a primitive root modulo $p^2$, for each $n\in\Z_{p^2}^*$, there exists a $0\le j_n<(p-1)p$ such that $n\equiv n_02^{j_n} \bmod {p^2}$. Then we have
$$
H_k(\beta^{n})=H_k(\beta^{n_0 2^{j_n}})=H_k(\beta^{n_0})^{2^{j_n}}=0,
$$
that is, all ($p^2-p$ many) elements $\beta^n$ for $n\in\Z_{p^2}^*$ are roots of $H_k(X)$. Hence we have
$$
\Phi(X)|H_k(X) ~~ \mathrm{in}~ \overline{\F}_2[X],
$$
where
$$
\Phi(X)=1+X^p+X^{2p}+\ldots+X^{(p-1)p}\in \F_2[X],
$$
the roots of which are exactly $\beta^n$ for $n\in\Z_{p^2}^*$.  Let
\begin{equation}\label{pi}
H_k(X)\equiv \Phi(X)\pi(X) \pmod {X^{p^2}-1}.
\end{equation}
Using the fact that
$$
X^p\Phi(X)  \equiv \Phi(X) \pmod {X^{p^2}-1},
$$
we restrict $\deg(\pi(X))<p$ and write
$$
\pi(X)=X^{v_0}+X^{v_1}+\ldots+X^{v_{t-1}} ~ \mathrm{with} ~ 0\le v_0<v_1<\ldots<v_{t-1}<p,
$$
where $t\ge 1$ since $H_k(X)$ is a nonzero polynomial. Then the exponent of each monomial in $\Phi(X)\pi(X) \bmod {X^{p^2}-1}$ forms the set
$$
\{v_j+lp : 0\le j\le t-1, 0\le l\le p-1\},
$$
which can be divided into two sets $A$ and $B$ with
$$
A=\{v_j+lp : 0\le j\le t-1, 0\le l\le p-1, v_j\neq 0, q_{p,w}(v_j+lp)\in \mathcal{I}\},
$$
$$
B=\{v_j+lp : 0\le j\le t-1, 0\le l\le p-1\}\setminus A.
$$
We note that by (\ref{addstruct}) $A$ contains $|A|$ many numbers with
$$
|A|=\left\{
\begin{array}{cl}
(t-1)|\mathcal{I}|, & \mathrm{if}~ v_0=0, \\
t|\mathcal{I}|, & \mathrm{otherwise},
\end{array}
\right.
$$
and $B$ contains $tp-|A|$ many numbers.

Hence, from (\ref{Hk}) and (\ref{pi}), we find that
the set of the exponents of monomials in $e(X)$ is
$$
( \cup_{l\in \mathcal{I}}D_l\setminus A)\cup B,
$$
the cardinality of which is
$$
(p-1)|\mathcal{I}|-|A|+|B|=(p-1)|\mathcal{I}|+tp-\left\{
\begin{array}{cl}
2(t-1)|\mathcal{I}|, & \mathrm{if}~ v_0=0, \\
2t|\mathcal{I}|, & \mathrm{otherwise}.
\end{array}
\right.
$$
That is, $k=(p-1)|\mathcal{I}|-|A|+|B|$ since $e(X)$ contains $k$ many terms. However, the
assumption of $1\le |\mathcal{I}|\le (p-1)/2$
implies $tp-2t|\mathcal{I}|>0$ and hence
$$
(p-1)|\mathcal{I}|-|A|+|B|>(p-1)|\mathcal{I}|>k,
$$
a contradiction.
So $H_k(\beta^{n})\neq 0$ for all $n\in \Z_{p^2}^*$.

 On the other hand, by Lemma \ref{lemma-add} we get
\begin{equation}\label{roots}
H_k(\beta^{ip})=e(\beta^{ip})+\left\{
\begin{array}{ll}
0, & \mathrm{if}~ i= 0,\\
|\mathcal{I}|, & \mathrm{if}~ 1\le i<p.
\end{array}
\right.
\end{equation}
Hence below we only need to consider the number of roots of the form $\beta^{ip}  ~ (0\le i<p)$ for $H_k(X)$.

First, if $|\mathcal{I}|$ is odd, for $k=0$ (in this case $e(X)$ will not occur) it is easy to see that
$$
LC^{\F_2}_0((h_u))=LC^{\F_2}((h_u))=p^2-1.
$$
For $1\le k<p-1$, we first consider $e(X)=X^{i_0 p}$ for some $0\le i_0<p$, i.e., $(h_u)$ is changed only one term at the position $i_0 p$ per period, we have $e(\beta^{ip})=1$ for all $0\le i<p$ and there are exactly $p-1$ many $\beta^{ip}  ~ (1\le i<p)$ such that $H_k(\beta^{ip})=0$. However, for any other $e(X)$ with $k~ (1\le k<p-1)$ terms, the number of such kind of roots of $H_k(X)$ does not increase, since  $e(X)$ satisfying
$$
e(\beta^{ip})=\left\{
\begin{array}{ll}
0, & \mathrm{if}~ i= 0,\\
1, & \mathrm{if}~ 1\le i<p,
\end{array}
\right.
$$
which guarantees all $\beta^{ip} ~(0\le i<p)$ are roots of $H_k(X)$, should be of the form
$$
e(X)\equiv  X+X^2+\ldots+X^{p-1} \pmod {X^p-1}
$$
by Lemma \ref{lemma-p}, in other words, $e(X)$ should contain at least $p-1$ many terms if the number of roots of  $H_k(X)$ increases (from $p-1$ to $p$). So we derive
$$
LC^{\F_2}_k((h_u))=LC^{\F_2}_1((h_u))=p^2-p+1  ~~\mathrm{for}~~ 1\le k<p-1.
$$
For $p-1\le k< (p-1)|\mathcal{I}|$ and $|\mathcal{I}|>1$, one can choose $e(X)$ with $p-1$ terms, as mentioned above,  of the form $e(X)\equiv  X+X^2+\ldots+X^{p-1} \bmod {X^p-1}$ such that all $\beta^{ip}$'s  $(0\le i<p)$ are roots of $H_k(X)$.  Hence, we get
$$
LC^{\F_2}_k((h_u))=LC^{\F_2}_{p-1}((h_u))=p^2-p  ~~\mathrm{for}~~ p-1\le k< (p-1)|\mathcal{I}|.
$$

Second, we turn to the case of even $|\mathcal{I}|$. When $k=0$, all $\beta^{ip}$ $(0\le i<p)$ are roots of  $H_k(X)$ from (\ref{roots}). No other possible roots of the form $\beta^{n}$ will occur for $1\le k< (p-1)|\mathcal{I}|$.
So we have
$$
LC^{\F_2}_k((h_u))=LC^{\F_2}_0((h_u))=p^2-p ~~\mathrm{for}~~ 0\le k< (p-1)|\mathcal{I}|.
$$
We complete the proof.   ~\hfill $\square$\\

In Theorem \ref{klc-2-primitive}, we restrict $1\le |\mathcal{I}|\le (p-1)/2$. For $|\mathcal{I}|> (p-1)/2$, we can similarly consider the complementary sequence, denoted by $(h'_u)$, of $(h_u)$, i.e., $h'_u\equiv h_u+1 \bmod 2$ for all $u\ge 0$. The difference between the ($k$-error) linear complexity of $(h_u)$ and that of  $(h'_u)$ is at most $1$ by the fact that
$$
\frac{H^c(X)+e(X)}{X^{p^2}-1}=\frac{H(X)+e(X)}{X^{p^2}-1}+\frac{1}{X-1},
$$
where $H^c(X)$ is the generating polynomial of $(h'_u)$, $H(X)$ is the generating polynomial of $(h_u)$ and $e(X)$ is the error polynomial. In this case, one might ask how about the $k$-error linear complexity for the complementary sequence $(h'_u)$, which in fact is defined by
$$
h'_u=\left\{
\begin{array}{ll}
1, & \mathrm{if}\,\ u\bmod {p^2}\in \cup_{l\in \mathcal{J}} D_l\cup P,\\
0, & \mathrm{otherwise},
\end{array}
\right. \quad u\ge 0,
$$
where $\mathcal{J}$ is a non-empty subset of $\{0,1,\ldots,p-1\}$ with $1\le |\mathcal{J}|\le (p-1)/2$.
(Note that $h'_u=1$ for $u\in P$, but $h_u=0 $ in this case.)  In particular, we can get some balanced binary sequences when $|\mathcal{J}|= (p-1)/2$ for certain special applications.

Fortunately, following the same way as the proof of Theorem \ref{klc-2-primitive}, we get
\[
 LC^{\F_2}_k((h'_u))=\left\{
\begin{array}{cl}
p^2-p+1, & \mathrm{if}\,\  0\le k<p-1, \\
p^2-p, & \mathrm{if}\,\ p-1\le k<(p-1)|\mathcal{J}|, \\
p, & \mathrm{if}\,\  k= (p-1)|\mathcal{J}|, \\
0, & \mathrm{if}\,\ k\ge (p-1)|\mathcal{J}|+1,
\end{array}
\right.\\
\]
if $|\mathcal{J}|$ is odd, and otherwise
\[
 LC^{\F_2}_k((h'_u))=\left\{
\begin{array}{cl}
p^2, & \mathrm{if}\,\ k=0, \\
p^2-p, & \mathrm{if}\,\ 1\le k<(p-1)|\mathcal{J}|, \\
p, & \mathrm{if}\,\  k= (p-1)|\mathcal{J}|, \\
0, & \mathrm{if}\,\ k\ge (p-1)|\mathcal{J}|+1,
\end{array}
\right.
\]
if $2$ is a primitive root modulo $p^2$.

The statement of the $k$-error linear complexities of $(e_u)$ and $(f_u)$ follows from Theorem \ref{klc-2-primitive} directly.
We describe it in the following corollary.

\begin{corollary}
Let $(e_u)$ and $(f_u)$ be the binary sequences of period $p^{2}$ defined in
(\ref{binarythreshold}) and (\ref{binarylegendre}), respectively. If $2$ is a primitive root modulo $p^2$, then
their $k$-error linear complexity  over $\F_2$ satisfies
\[
 LC^{\F_2}_k((e_u))=LC^{\F_2}_k((f_u))=\left\{
\begin{array}{cl}
p^2-1, & \mathrm{if}\,\ k=0, \\
p^2-p+1, & \mathrm{if}\,\ 1\le k<p-1, \\
p^2-p, & \mathrm{if}\,\ p-1\le k< (p-1)^2/2, \\
0, & \mathrm{if}\,\ k\ge (p-1)^2/2,
\end{array}
\right.
\]
if $p \equiv 3 \bmod 4$, and otherwise
\[
LC^{\F_2}_k((e_u))=LC^{\F_2}_k((f_u))=\left\{
\begin{array}{cl}
p^2-p, & \mathrm{if}\,\ 0\le k<(p-1)^2/2, \\
0, & \mathrm{if}\,\ k\ge (p-1)^2/2.
\end{array}
\right.\\
\]
\end{corollary}

For $w=1$, the result is somewhat different because of (\ref{value-2}) and we present it in the following separate
theorem.

\begin{theorem}\label{klc-2-primitive-w=1}
Let $(h_u)$ be the binary sequence of period $p^{2}$ defined in (\ref{hhhh-w=1}) using polynomial quotients (\ref{poly-def}) with $w=1$ and a non-empty subset $\mathcal{I}$ of $\{0,1,\ldots,p-1\}$ with $1\le |\mathcal{I}|\le (p-1)/2$. If $2$ is a primitive root modulo $p^2$, then
the $k$-error linear complexity  over $\F_2$ of $(h_u)$  satisfies
\[
 LC^{\F_2}_k((h_u))=\left\{
\begin{array}{cl}
p^2-p+1, & \mathrm{if}\,\ 0\le k<p, \\
p^2-p, & \mathrm{if}\,\ p\le k< p|\mathcal{I}|, |\mathcal{I}|>1,\\
0, & \mathrm{if}\,\ k\ge p|\mathcal{I}|,
\end{array}
\right.
\]
if $|\mathcal{I}|$ is odd, and otherwise
\[
 LC^{\F_2}_k((h_u))=\left\{
\begin{array}{cl}
p^2-p, & \mathrm{if}\,\ 0\le k<p|\mathcal{I}|, \\
0, & \mathrm{if}\,\ k\ge p|\mathcal{I}|.
\end{array}
\right.
\]
\end{theorem}
Proof. The proof is similar to that of Theorem \ref{klc-2-primitive}. Here we present a sketch. Let
$$
H_k(X)=\sum\limits_{l\in \mathcal{I}}D_{l}(X)+\sum\limits_{l\in \mathcal{I}}X^{lp}+e(X)\in \F_2[X]
$$
be the generating polynomial of the sequence obtained from $(h_u)$ by changing exactly $k$ terms of $(h_u)$ per period,
where $e(X)$ is the corresponding error polynomial with $k$ many  monomials.

For $k< p|\mathcal{I}|$, under the assumption of $2$ being primitive root modulo $p^2$, we can show $H_k(\beta^{n})\neq 0$ for all $n\in \Z_{p^2}^*$, as proved in  Theorem \ref{klc-2-primitive}. So we only need to determine the number of roots of the form $\beta^{ip}  ~ (0\le i<p)$ for $H_k(X)$.
By Lemma \ref{lemma-add}, we have
\begin{eqnarray*}
H_k(\beta^{ip})& = & \sum\limits_{l\in \mathcal{I}}D_{l}(\beta^{ip})+\sum\limits_{l\in \mathcal{I}}(\beta^{ip})^{lp}+e(\beta^{ip})\\
& =&e(\beta^{ip})+ \left\{
\begin{array}{ll}
|\mathcal{I}|, & \mathrm{if}~ i= 0,\\
0, & \mathrm{if}~ 1\le i<p.
\end{array}
\right.
\end{eqnarray*}
For odd $|\mathcal{I}|$, all $\beta^{ip}  ~ (1\le i<p)$ are roots of $H_k(X)$ when $k=0$ and
$H_k(X)$ has one more root if $e(X)$ satisfies
$$
e(\beta^{ip})=\left\{
\begin{array}{ll}
1, & \mathrm{if}~ i= 0,\\
0, & \mathrm{if}~ 1\le i<p,
\end{array}
\right.
$$
from which we derive by Lemma \ref{lemma-p}
$$
e(X)\equiv 1+X+X^2+\ldots+X^{p-1} \pmod {X^p-1}.
$$
That is to say, only that $e(X)$ modulo $X^p-1$ is of the form above, which contains $p$ terms, can guarantee that all $\beta^{ip} ~(0\le i<p)$ are roots of $H_k(X)$,
thus
$$
LC^{\F_2}_k((h_u))=LC^{\F_2}_0((h_u))=p^2-p+1 ~~\mathrm{for}~~ 0\le k< p
$$
and
$$
LC^{\F_2}_k((h_u))=LC^{\F_2}_p((h_u))=p^2-p ~~\mathrm{for}~~ p\le k< p|\mathcal{I}|.
$$
For even $|\mathcal{I}|$, all $\beta^{ip}  ~ (0\le i<p)$ are roots of $H_0(X)$  and any $e(X)$ with $k$ terms for  $k< p|\mathcal{I}|$ will not increase
the number of the common roots of $H_k(X)$ and $X^{p^2}-1$. Then the result follows. ~\hfill $\square$\\

We restrict that $2$ is a primitive root modulo $p^2$ in the theorems above. A
conjecture of Artin suggests that approximately $3/8$ of all primes
have $2$ as a primitive element (\cite[p.81]{Shanks}), and it is
very seldom that a primitive element modulo the prime $p$ is not
 a primitive element modulo $p^2$.
If $2$ is not a primitive root modulo $p^2$, it seems that our method is not suitable for computing the exact number of the common roots of  $H_k(X)$ and $X^{p^2}-1$ without additional ideas, as mentioned in the proof of Theorem \ref{klc-2-primitive}. But we have some partial results, as described in the following theorem, under certain special conditions. We conjecture that Theorems \ref{klc-2-primitive} and \ref{klc-2-primitive-w=1} are true for most primes $p$, e.g. $p$ satisfying $2^{p-1}
\not\equiv 1 \pmod {p^2}$, see \cite{CDP1997,CD} for the applications of such primes. We note that $2^{p-1}
\not\equiv 1 \pmod {p^2}$ if and only if the order of $2$ modulo $p^2$ is lager than $p$.

\begin{theorem}\label{klc-2-general}
Let $\mathcal{I}\subseteq \{0,1,\ldots,p-1\}$ with $1\le |\mathcal{I}|\le (p-1)/2$ and the order of $2$ modulo $p^2$ be $\lambda p$ with $1<\lambda\le p-1$ and $\lambda|(p-1)$.

(i). Let $(h_u)$ be the binary sequence of period $p^{2}$ defined in (\ref{hhhh}) using polynomial quotients (\ref{poly-def}) with $w\ge 2$ and $\mathcal{I}$.

(ii). Let $(h_u)$ be the binary sequence of period $p^{2}$ defined in (\ref{hhhh-w=1})  using polynomial quotients (\ref{poly-def}) with $w=1$  and $\mathcal{I}$.

If $0\le k<(p-1)|\mathcal{I}|$ for (i) or $0\le k<p|\mathcal{I}|$ for (ii), the $k$-error linear complexity over $\F_2$ of $(h_u)$  satisfies
\[
 LC^{\F_2}_k((h_u))\ge \lambda p,
 \]
 and otherwise $LC^{\F_2}_k((h_u))=0$.
\end{theorem}
Proof. First for $0\le k<(p-1)|\mathcal{I}|$ for (i), according to the proof of Theorem \ref{klc-2-primitive}, there do exist an $n_0\in \Z_{p^2}^*$ such that
 $H_k(\beta^{n_0})\neq 0$ for $H_k(X)$, the generating polynomial of the sequence obtained from $(h_u)$ by changing exactly $k$ terms of $(h_u)$ per period. (Otherwise, we will get a more accurate result, as described in Theorem \ref{klc-2-primitive}.)
Thus there are at least $\lambda p$ many $n\in \{n_02^j \bmod {p^2} : 0\le j<\lambda p\}$ such that $H_k(\beta^{n})\neq 0$. Then the result follows.

For the case of  $0\le k<p|\mathcal{I}|$ for (ii), the discussion is similar by using  the proof of Theorem \ref{klc-2-primitive-w=1}. ~\hfill $\square$

\section{$k$-Error Linear Complexity over $\F_p$}\label{LC-p}

In this section, we view the binary sequences $(h_u)$ defined in  (\ref{hhhh}) and (\ref{hhhh-w=1}) as sequences over $\F_p$
and consider their ($k$-error) linear complexity over $\F_p$, which is also an interesting problem for binary sequences. Such kind of work has been done in many references, such as \cite{AM,AMW,AW06,AW,BW,CY,ESK,GLSW,HKN,HMMS}.

We will employ the $j$-th Hasse derivative of a polynomial $F(x)=a_0+a_1X+\ldots+a_{T-1}X^{T-1}\in\F_p[X]$, which is defined to be
$$
F^{(j)}(X)=\sum\limits_{n=j}^{T-1} \binom{n}{j}a_iX^{n-j}, ~~ j\ge 1.
$$
The multiplicity of $\mu$ as a root of $F(X)$ is $n$ if $F(\mu)=F^{(1)}(\mu)=\ldots =F^{(n-1)}(\mu)=0$ and $F^{(n)}(\mu)\neq 0$, see e.g. \cite[Ch 6.4]{LN} for details.

Before presenting the main results, we introduce a technical lemma, which will be used in the proofs.

\begin{lemma}\label{D-derivative}
With notations of $D_l~(0\le l<p)$ defined in Section \ref{intro}. Let $D_l(X)= \sum\limits_{u\in D_l}X^u \in \mathbb{F}_p[X]$
and $D^{(j)}_l(X)$ be the $j$-th Hasse derivative of $D_l(X)$ for $0\le l<p$. Then for $0\le l<p$, we have
$$
D_l(X)\equiv  (X-1)^{p-1}-1 \pmod {X^p-1}
$$
and hence
$$
D_l(1)=p-1,~ D^{(j)}_l(1)=0 ~ \mathrm{and}~ D^{(p-1)}_l(1)=1,
$$
where $1\le j\le p-2$.
\end{lemma}
Proof. In the proof of Lemma \ref{lemma-add}, we have shown that
$$
\{u\pmod p: u\in D_l\}=\Z_p^*, ~~~0\leq l < p,
$$
i.e.,
$$
D_l=\{v+m_{lv}p : 1\le v<p,  m_{lv}=(wv^{w-1})^{-1}(l-q_{p,w}(v))\bmod p \},
$$
from which we derive
\begin{eqnarray*}
D_l(X)& \equiv & X+X^{2}+\ldots+X^{p-1} \\
      & \equiv & (1+X+X^{2}+\ldots+X^{p-1})-1\\
      & \equiv & \frac{X^{p}-1}{X-1}-1\\
       & \equiv & \frac{(X-1)^{p}}{X-1}-1\\
       & \equiv & (X-1)^{p-1}-1  \pmod {X^p-1}.
\end{eqnarray*}
Then write
$$
D_l(X)= (X-1)^{p-1}-1 +\eta(X)(X^p-1) \in \F_p[X]
$$
for some $\eta(X) \in \F_p[X]$, it is  easy to check the rest equalities by using
$$
D^{(j)}_l(X)=\binom{p-1}{j}(X-1)^{p-1-j}+\sum\limits_{\stackrel{j_1+j_2=j}{0\le j_1,j_2\le j}}\binom{p}{j_1}(X-1)^{p-j_1}\eta^{(j_2)}(X)
$$
for $1\leq j < p$, where we use $X^p-1=(X-1)^p$. ~\hfill $\square$

Now we present our main results.

\begin{theorem}\label{klc-p-w=1}
Let $(h_u)$ be the (binary) sequence of period $p^{2}$ defined in (\ref{hhhh-w=1}) using polynomial quotients (\ref{poly-def}) with $w=1$ and a non-empty subset $\mathcal{I}$ of $\{0,1,\ldots,p-1\}$ with $1\le |\mathcal{I}|\le (p-1)/2$.
Then we have
$$
LC^{\F_p}_k((h_u))=LC^{\F_p}_0((h_u))=p^2-p+1
$$
for $0\le k<p$, and $LC^{\F_p}_k((h_u))\le p^2-p$ for $k\ge p$.
\end{theorem}
Proof. Let
$$
H_k(X)=\sum\limits_{l\in \mathcal{I}}(D_{l}(X)+X^{lp})+e(X)\in \F_p[X]
$$
be the generating polynomial of the sequence obtained from $(h_u)$ by changing exactly $k$ terms of $(h_u)$ per period,
where $e(X)$ is the corresponding error polynomial with $k$ many  monomials. In particular, $H_0(X)=\sum\limits_{l\in \mathcal{I}}(D_{l}(X)+X^{lp})$ is the generating polynomial of $(h_u)$. Since $X^{p^2}-1=(X-1)^{p^2}$ over $\F_p$, we only need to consider the multiplicity of $1$ as a root of $H_k(X)$.

It is easy to check by Lemma \ref{D-derivative} that
$$
H_0(1)=0, ~ H_0^{(j)}(1)=0 ~\mathrm{for}~1\le j\le p-2, ~
H_0^{(p-1)}(1)=|\mathcal{I}|\neq 0,
$$
where $H^{(j)}_0(X)$ is the $j$-th Hasse derivative of $H_0(X)$. So we have
$$
(X-1)^{p-1}\| H_0(X),
$$
where the notation `$\|$' means $(X-1)^{p-1}\mid H_0(X)$ but $(X-1)^{p}\nmid  H_0(X)$.
Hence the linear complexity of $(h_u)$ is
$$
LC^{\F_p}((h_u))=LC^{\F_p}_0((h_u))=p^2-(p-1)
$$
by (\ref{licom}).

Now we consider the case of $k\ge 1$. For $e(X)$ with $k$ terms, since $(X-1)^{p-1}\| H_0(X)$ it is easy to see that
$$
(X-1)^{p-1}\| H_k(X)
$$
if $(X-1)^{p}|e(X)$, and
$$
(X-1)^{m}\| H_k(X)
$$
if  $(X-1)^{m}\| e(X)$ and $m\le p-2$. So for such $e(X)$, the multiplicity of $1$ as a root of $H_k(X)$ is at most $p-1$ and hence the $k$-error linear complexity will not decrease. Now
we assume $(X-1)^{p-1}\| e(X)$ and write
$$
e(X)\equiv (X-1)^{p-1}(\alpha_0X^{v_0}+\alpha_1X^{v_1}+\ldots+\alpha_{t-1}X^{v_{t-1}})  \pmod {X^p-1},
$$
where $\alpha_{0},\ldots,\alpha_{t-1}\in\F_p^*$, $\alpha_{0}+\ldots+\alpha_{t-1}\neq 0$ and $0\le v_0<v_1<\ldots<v_{t-1}\le p^2-p$.
Using the facts that
$$
(X-1)^{p-1}=\frac{X^p-1}{X-1}=1+X+X^{2}+\ldots+X^{p-1}\in\F_p[X]
$$
and
\begin{eqnarray*}
&       &(1+X+X^{2}+\ldots+X^{p-1})\cdot \alpha_iX^{v_i}\\
&\equiv &(1+X+X^{2}+\ldots+X^{p-1})\cdot (\alpha_iX^{v_i}-\alpha_iX^{v_i-1}+\alpha_iX^{v_i-1})\\
&\equiv &(1+X+X^{2}+\ldots+X^{p-1})\cdot \alpha_iX^{v_i-1}\\
&\equiv &\ldots\\
&\equiv &(1+X+X^{2}+\ldots+X^{p-1})\cdot \alpha_i  \pmod {X^p-1},
\end{eqnarray*}
we get
$$
e(X)\equiv (1+X+X^{2}+\ldots+X^{p-1})(\alpha_0+\alpha_1+\ldots+\alpha_{t-1})  \pmod {X^p-1}.
$$
That is to say, $e(X)$ modulo $X^{p}-1$ should be of the form above and it has at least $p$ terms if $(X-1)^{p-1}\| e(X)$.

Hence we conclude that, if $1\le k<p$, the multiplicity of $1$ as a root of $e(X)$, which contains $k$ terms, is not equal to $p-1$. (Otherwise, $e(X)$ modulo $X^{p}-1$ has at most $k$ terms, a contradiction.) So we have
$$
(X-1)^{p}\nmid H_k(X) ~\mathrm{for}~ 1\le k<p
$$
and we derive the desired result.

For $k=p$, one can choose
$$
e(x)=-\alpha(1+X+X^{2}+\ldots+X^{p-1})=-\alpha(X-1)^{p-1}\in\F_p[X],
$$
where $\alpha =\frac{H_0(X)}{(X-1)^{p-1}}\bigg|_{X=1}\neq 0$. From
$$
H_k(X)=H_0(X)+e(X)=(X-1)^{p-1}\left(\frac{H_0(X)}{(X-1)^{p-1}}-\alpha\right),
$$
we find $(X-1)^{p}\mid H_k(X)$ and the value $LC^{\F_p}_p((h_u))\le p^2-p$.     ~\hfill $\square$

\begin{theorem}\label{klc-p-w=2}
Let $(h_u)$ be the (binary) sequence of period $p^{2}$ defined in (\ref{hhhh}) using polynomial quotients (\ref{poly-def}) with $2\le w\le p-1$ and a non-empty subset $\mathcal{I}$ of $\{0,1,\ldots,p-1\}$ with $1\le |\mathcal{I}|\le (p-1)/2$.
Then we have
\[
 LC^{\F_p}_k((h_u))=\left\{
\begin{array}{cl}
p^2, & \mathrm{if}\,\ k=0, \\
p^2-p+1, & \mathrm{if}\,\ 1\le k<p-1,
\end{array}
\right.
\]
and $LC^{\F_p}_k((h_u))\le  p^2-p$ for $k\ge p-1$.
\end{theorem}
Proof. Let
$$
H_k(X)=\sum\limits_{l\in \mathcal{I}}D_{l}(X)+e(X)\in \F_p[X]
$$
be the generating polynomial of the sequence obtained from $(h_u)$ by changing exactly $k$ terms of $(h_u)$ per period,
where $e(X)$ is the corresponding error polynomial with $k$ many  monomials.

We check that $H_0(1)=(p-1)|\mathcal{I}|\neq 0 \bmod p$, hence $LC^{\F_p}((h_u))=p^2$ by (\ref{licom}).

Below we consider the case of $k=1$. For $e(X)=\xi X^{ip}$ for any $\xi\in\F_p^*$ and $0\le i<p$,
we have
$$
H_1(1)=(p-1)|\mathcal{I}|+\xi \left\{
\begin{array}{cl}
=0, & \mathrm{if}~\xi=|\mathcal{I}|,\\
\neq 0, & \mathrm{otherwise}
\end{array}
\right.
$$
 and
\begin{eqnarray*}
H^{(j)}_1(X)\bigg|_{X=1}& = & \sum\limits_{l\in \mathcal{I}}D^{(j)}_{l}(X)\bigg|_{X=1}+\binom{ip}{j}\xi X^{ip-j}\bigg|_{X=1}\\
                        & = &
\left\{
\begin{array}{cl}
0, & \mathrm{if}~1\le j\le p-2,\\
|\mathcal{I}|, & \mathrm{if}~ j=p-1
\end{array}
\right.
\end{eqnarray*}
by Lemma \ref{D-derivative}. So we derive
$$
(X-1)^{p-1}\| H_1(X) ~~\mathrm{if} ~~ \xi=|\mathcal{I}|
$$
and
$$
(X-1)\nmid H_1(X) ~~\mathrm{if} ~~ \xi\neq |\mathcal{I}|.
$$
For $e(X)=\xi X^{n}$ for any $\xi\in\F_p^*$ and $n\in\Z_{p^2}^*$,  we find that $H_1(1)=(p-1)|\mathcal{I}|+\xi$ and
$$
H^{(1)}_1(X)\bigg|_{X=1}=\sum\limits_{l\in \mathcal{I}}D^{(1)}_{l}(X)\bigg|_{X=1}+n\xi X^{n-1}\bigg|_{X=1}=n\xi \neq 0,
$$
hence the multiplicity of $1$ as a root of $ H_1(X)$ is $\le 1$. So we conclude that
$$LC^{\F_p}_1((h_u))=p^2-(p-1).$$

Now we want to find the smallest $k\ge 2$ such that
$$
(X-1)^{p}| H_k(X).
$$
From $H_k(1)=0$ and $H^{(j)}_k(1)=0$  for $1\le j\le p-1$, we have
$$
\left\{
\begin{array}{l}
e(1)=|\mathcal{I}|,\\
e^{(j)}(1)=0 ~\mathrm{for}~1\le j\le p-2,\\
e^{(p-1)}(1)=(p-1)|\mathcal{I}|,
\end{array}
\right.
$$
by Lemma \ref{D-derivative}, where  $e^{(j)}(X)$ is the $j$-th Hasse derivative of $e(X)$. We define a new polynomial $\widetilde{e}(X)\in\F_p[X]$ with
$$
\widetilde{e}(X)=e(X)+(p-|\mathcal{I}|)X^{ip}
$$
for some $0\le i<p$, and compute
$$
\widetilde{e}(1)=0, ~ \widetilde{e}^{(j)}(1)=0 ~\mathrm{for}~1\le j\le p-2, ~
\widetilde{e}^{(p-1)}(1)=(p-1)|\mathcal{I}|\neq 0.
$$
Then we have $(X-1)^{p-1}\| \widetilde{e}(X)$. Following the proof of Theorem \ref{klc-p-w=1}, we derive
$$
\widetilde{e}(X)\equiv \mu (1+X+X^{2}+\ldots+X^{p-1})\pmod {X^p-1}
$$
for $\mu\in\F_p^*$. Hence $e(X)$ should be of the form
$$
e(X)=\widetilde{e}(X)+|\mathcal{I}|X^{ip}\equiv |\mathcal{I}|+\mu (1+X+X^{2}+\ldots+X^{p-1}) \pmod {X^p-1},
$$
which contains at least $p-1$ terms, since $|\mathcal{I}|+\mu$ can take $0$ as its output. Hence $k\ge p-1$.

So we conclude that if $1\le k<p-1$, $(X-1)^{p}\nmid  H_k(X)$, from which the first desired result follows.
For $k=p-1$, one can directly choose
$$
e(x)=-|\mathcal{I}|(X+X^{2}+\ldots+X^{p-1})=|\mathcal{I}|-|\mathcal{I}|(X-1)^{p-1}\in\F_p[X],
$$
and then compute
$$
H_k(1)=H_0(1)-(p-1)|\mathcal{I}|=0
$$
and
$$
H^{(j)}_k(X)\bigg|_{X=1}= H^{(j)}_0(X)\bigg|_{X=1}- |\mathcal{I}|\binom{p-1}{j}(X-1)^{p-1-j}\bigg|_{X=1}=0
$$
for $1\le j\le p-1$ by Lemma \ref{D-derivative}, so we have $(X-1)^{p}\mid H_k(X)$ and  $LC^{\F_p}_{p-1}((h_u))\le p^2-p$.
 ~\hfill $\square$\\

It seems difficult for us to consider the case of larger $k$ without additional ideas. We leave it open. However, motivated by \cite{AMW,AW06}, we have a more accurate upper bound for $\mathcal{I}$ being the set of quadratic non-residues modulo $p$.
In this case, $(h_u)$  in (\ref{hhhh}) or (\ref{hhhh-w=1}) is in fact $(f_u)$ defined in (\ref{binarylegendre}).

Since
\begin{eqnarray*}
 q_{p,w}(u) & \equiv & q_{p,w}(i_0+i_1p)\\
            & \equiv & \sum\limits_{c=0}^{p-1}q_{p,w}(c)\left(1-(i_0-c)^{p-1} \right)+wi_0^{w-1}i_1 \pmod p,
\end{eqnarray*}
for all integers $u\equiv i_0+i_1p \pmod {p^2}$ with $0\le i_0, i_1<p$, according to \cite{AW06} we see that
$(f_u)$ can be represented by
$$
f_{i_0+i_1p+jp^2}=\rho(i_0,i_1) ~\mathrm{for~ all~ integers}~ 0\le i_0, i_1<p ~\mathrm{and}~ j,
$$
where the multivariate polynomial $\rho(X_0,X_1)\in \F_p[X_0,X_1]/\langle X^p_0-X_0, X^p_1-X_1\rangle$ is of the form
\begin{eqnarray*}
\rho(X_0,X_1) & = & 2^{-1}\left(\sum\limits_{c=0}^{p-1}q_{p,w}(c)\left(1-(X_0-c)^{p-1} \right)+wX_0^{w-1}X_1 \right)^{p-1}\\
&& -2^{-1}\left(\sum\limits_{c=0}^{p-1}q_{p,w}(c)\left(1-(X_0-c)^{p-1} \right)+wX_0^{w-1}X_1 \right)^{\frac{p-1}{2}}.
\end{eqnarray*}
We reduce $\rho(X_0,X_1)$ modulo $X^p_0-X_0$ and $X^p_1-X_1$ such that the degree strictly less than $p$ in each indeterminate.
And then the linear complexity over $\F_p$ of $(f_u)$ equals to $1+\deg(\rho(X_0,X_1))$, we refer the reader to \cite[Theorem 8]{BEP} for the assertion and the definition of the \emph{degree} of multivariate polynomials.

For $2\le w<p$, Substituting $0$ by $2^{-1}(=\frac{p+1}{2})$ at those positions $u$ with $u\bmod {p^2}\in D_0\cup P$ in $(f_u)$,
we get a new sequence $(\overline{f}_u)$ represented by the polynomial
$$
2^{-1}-2^{-1}\left(\sum\limits_{c=0}^{p-1}q_{p,w}(c)\left(1-(X_0-c)^{p-1} \right)+wX_0^{w-1}X_1 \right)^{\frac{p-1}{2}},
$$
from which we derive after some simple calculations
$$
LC^{\F_p}((\overline{f}_u))= \left\{
\begin{array}{ll}
(p-1)p/2+p, & \mathrm{if}~ w\equiv 1 \bmod 2,\\
(p-1)p/2+(p-1)/2+1, & \mathrm{otherwise},
\end{array}
\right.
$$
by \cite[Theorem 8]{BEP}. Since $|D_0\cup P|=2p-1$, we obtain an upper bound on the $k$-error linear complexity
of  $(f_u)$ defined in (\ref{binarylegendre}) as follows
$$
LC^{\F_p}_{k}((f_u))\le \left\{
\begin{array}{ll}
(p-1)p/2+p, & \mathrm{if}~ w\equiv 1 \bmod 2,\\
(p-1)p/2+(p-1)/2+1, & \mathrm{otherwise},
\end{array}
\right.
$$
for $k\ge 2p-1$.

For $w=1$, We only use $\{0\}$ instead of $P$ above and obtain
$$
LC^{\F_p}_k((f_u))\le  (p-1)p/2+1
$$
for $k\ge p$.\\

Finally, we mention a lower bound on the $k$-error linear complexity over $\F_p$ of $(h_u)$
defined in (\ref{hhhh})  or (\ref{hhhh-w=1}). From \cite[Theorem 8]{BEP}, each $p^2$-periodic sequence over $\F_p$ can be represented by a unique polynomial
$\varrho(X_0,X_1) \in \F_p[X_0,X_1]/\langle X^p_0-X_0, X^p_1-X_1\rangle$
with $\deg_{X_1}(\varrho(X_0,X_1))\ge 1$, otherwise the period is reduced to $p$. We find by (\ref{addstruct}) that
changing at most $k$ (smaller than the weight of $(h_u)$ per-period) terms from  $(h_u)$ will not reduce the period, hence
the $k$-error linear complexity over $\F_p$ is $\ge p+1$.

\section{Concluding Remarks}

In this paper, we study the \emph{error linear complexity spectrum} (see \cite{EKKLP} for details) of
$p^2$-periodic
binary sequences defined from the polynomial quotients, that is, we determine exact values of
their $k$-error linear complexity over the finite field $\F_2$
for all integers $k$ under the assumption of $2$ being a primitive root modulo $p^2$. Main results can be described in the following figures, which visually reflect how the linear complexity of the binary sequences decreases as the number $k$ of allowed bit changes increases. It is of interest to consider this problem for the case of $2$ being not a primitive root modulo $p^2$. We only estimate a lower bound on their $k$-error linear complexity if $2^{p-1} \not\equiv 1 \pmod {p^2}$, with which most primes $p$ are satisfied, see \cite{CDP1997}.

\begin{figure}[H]
\centering
\includegraphics[width=110mm,height=60mm]{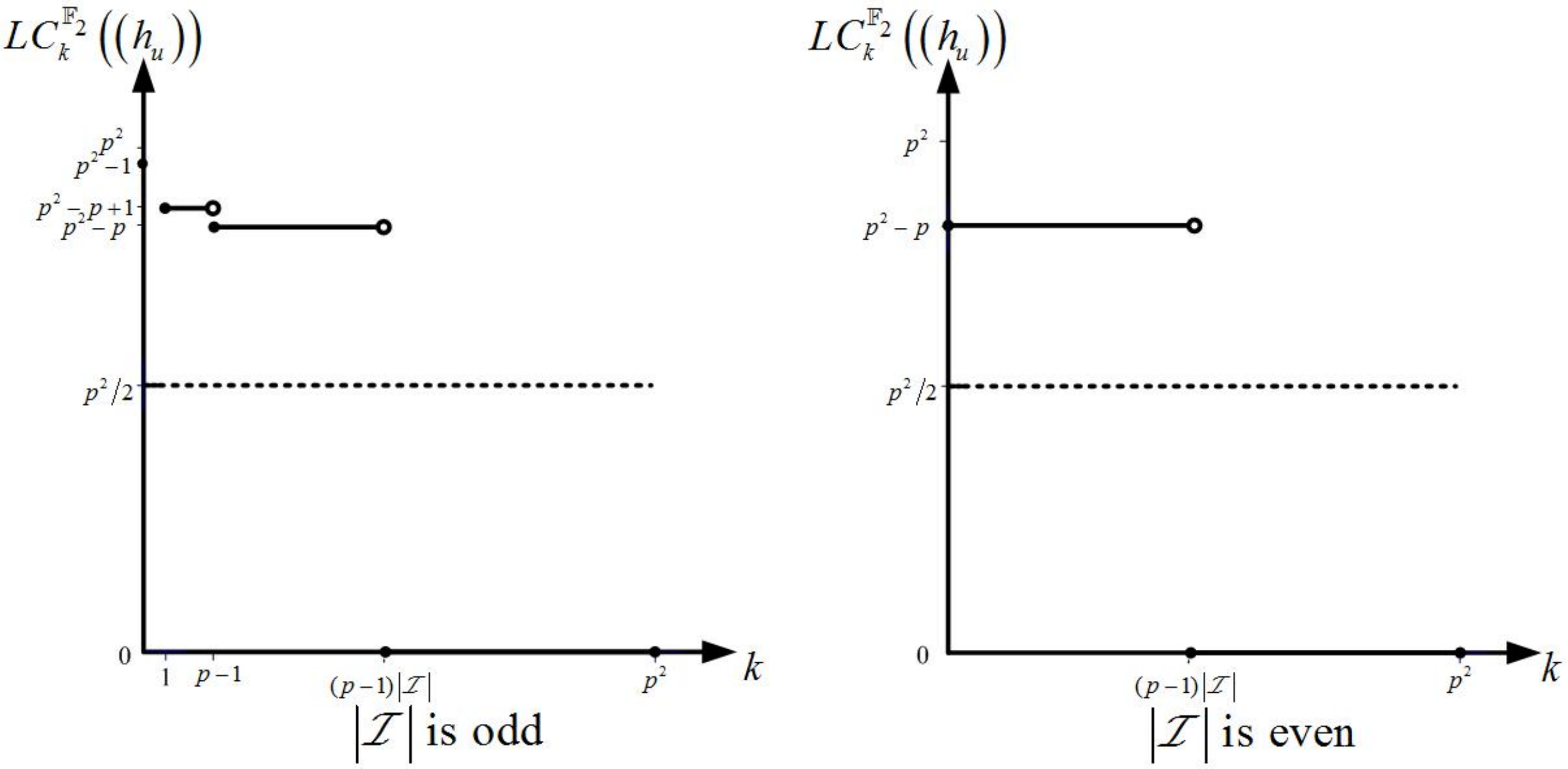}
\caption{Error linear complexity spectrum of $(h_u)$ when $2\le w<p$ (Theorem \ref{klc-2-primitive})}
\end{figure}

\begin{figure}[H]
\centering
\includegraphics[width=110mm,height=60mm]{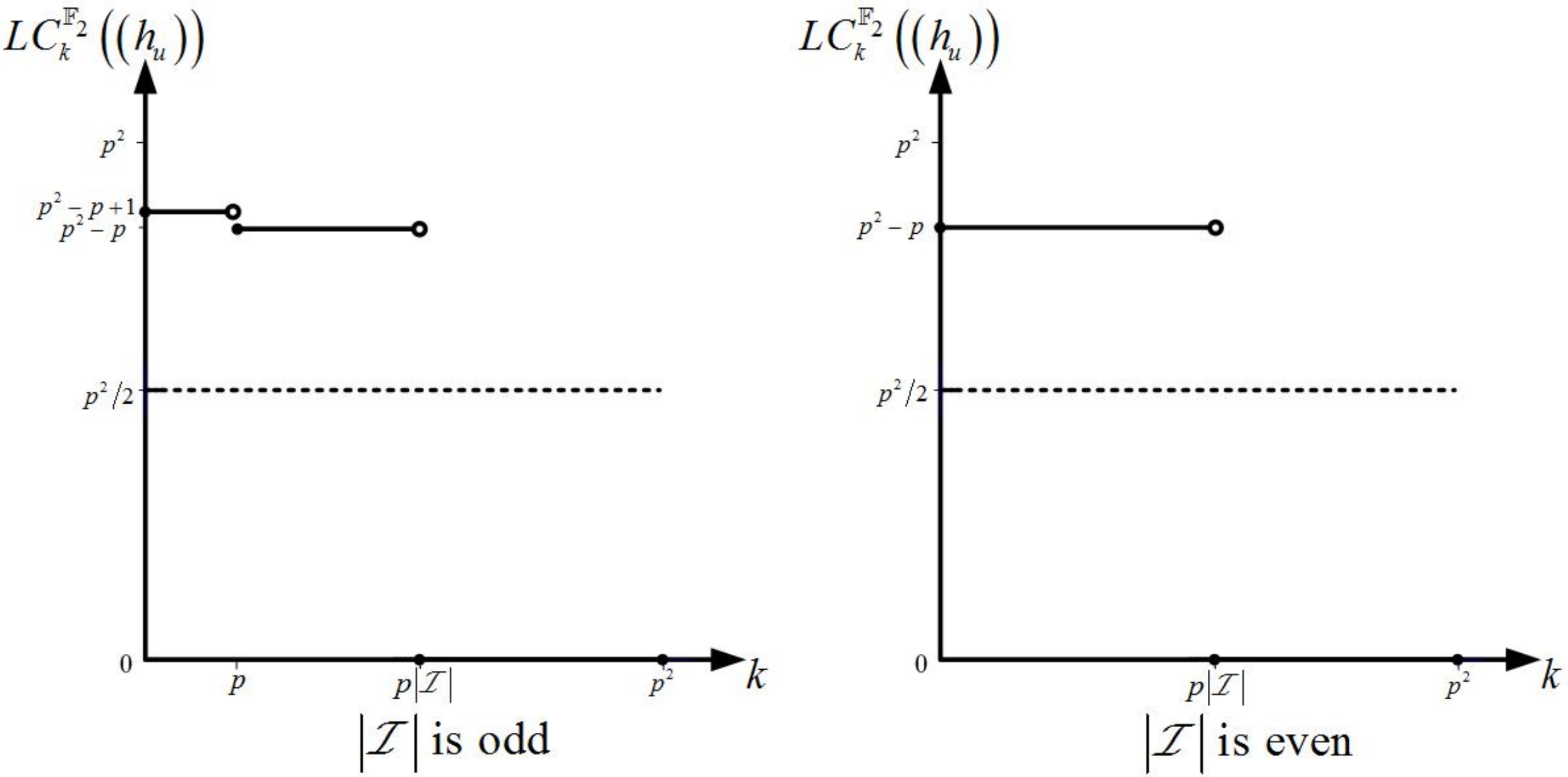}
\caption{Error linear complexity spectrum of $(h_u)$ when $w=1$ (Theorem \ref{klc-2-primitive-w=1})}
\end{figure}

We also view the binary sequences as sequences over the finite field $\F_p$
and determine their $k$-error linear complexity over $\F_p$ for either $0\le k<p$ when $w=1$ or $0\le k<p-1$ when $2\le w<p$. Results indicate that the linear complexity is large (close to the period) and  not significantly reduced by changing a few terms. It is interesting to consider this problem for larger $k$.

We finally remark that, the definition of binary sequences studied in this manuscript is related to
generalized cyclotomic classes modulo $p^2$, as you can see in Section \ref{intro}. In particular, the Fermat quotient
 $q_p(-)$ defines a group epimorphism from $\Z_{p^2}^*$ to $\Z_p$ by the fact, see e.g. \cite{OS}, that
$$
q_p(uv) \equiv q_p(u) + q_p(v) \pmod p, ~~\gcd(uv,p)=1.
$$
So if $g$ is a (fixed) primitive root modulo $p^2$, we have
$$
D_0=\{g^{jp} \bmod {p^2} : 0\le j<p\}
$$
and
$$
 D_{l\delta}=g^jD_0=\{g^{jp+l} \bmod {p^2} : 0\le j<p\},~~ 1\le l<p,
$$
where $\delta =q_p(g)$  and the subscript of $D$ is performed modulo $p$. (Note that for $w\neq p-1$, we don't have this property.) Sequences related to cyclotomic classes modulo a prime and generalized cyclotomic classes modulo the product of two
distinct primes have been widely investigated since several decades ago, the well-known basic examples are the Legendre sequences and the Jacobi sequences, see \cite{CDR,D97,D98,DHS} and references therein. As we know, the $k$-error linear complexity of the Jacobi sequences and their generalizations \cite{D97,D98} has not been solved thoroughly.
Hence we hope that our idea and method might be helpful for considering this problem and lead to furtherly study applications of the theory of cyclotomy in cryptography.

\section*{Acknowledgements}

The authors wish  to thank Arne Winterhof for helpful suggestions.


Z.X.C. was partially supported by the National Natural Science
Foundation of China under grant No. 61170246 and the Special Scientific Research Program in Fujian Province Universities of China under grant No. JK2013044.

Z.H.N. was partially supported by the Shanghai Leading Academic Discipline Project under grant No. J50103,
the National Natural Science Foundation of China  under grants No. 61074135, 61272096 and 61202395, the Program for New Century Excellent Talents in University  under grant NCET-12-0620, and the Shanghai Municipal Education Commission Innovation Project.

C.H.W. was partially supported by the Foundation item of the Education Department of Fujian Province of China under grants No. JA12291 and JB12179.

Parts of this paper were written during a very pleasant visit of the
first author to RICAM in Linz. He wishes to thank for the hospitality.

\end{document}